\documentclass[twocolumn]{openjournal}
\usepackage{graphicx}
\usepackage{accsupp}
\usepackage{hyperref}
\usepackage{ragged2e}
\usepackage{booktabs}
\hypersetup{breaklinks,colorlinks,citecolor=blue,urlcolor=cyan}
\usepackage{bm}
\renewcommand{\vec}[1]{\bm{#1}}

\newcommand{\orcidauthor}[3]{\author{#2$^{#3}$ \href{http://orcid.org/#1}{\includegraphics[scale=0.04]{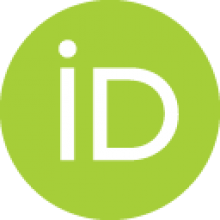}}}}
\begin{document}

\title{\vspace{-6mm}Galaxy mergers and disk angular momentum evolution: stellar halos as a critical test\vspace{-17mm}}

\orcidauthor{0000-0002-5564-9873}{Eric F.\ Bell$^1$ }{\star}
\orcidauthor{0000-0001-9269-8167}{Richard D'Souza$^2$ }{}
\orcidauthor{0000-0002-6257-2341}{Monica Valluri$^1$ }{}
\orcidauthor{0000-0003-2294-4187}{Katya Gozman$^1$ }{}
\affil{$^1$ Department of Astronomy, University of Michigan, 1085 S. University Ave, Ann Arbor, MI 48109-1107, USA \\
$^2$ Vatican Observatory, Specola Vaticana, V-00120, Vatican City State}

\thanks{$^\star$E-mail: \href{mailto:ericbell@umich.edu}{ericbell@umich.edu}}

\begin{abstract}

We investigate the role of hierarchical assembly in the angular momentum (AM) evolution of galaxies using a sample of 471 Milky Way-mass galaxies from the TNG-50 simulation. While galaxy orientation is often attributed to tidal torques and the cooling of gas within halos, we demonstrate that galaxy reorientation (tilting) is a common consequence of satellite accretion. Specifically, 80$\pm2$\% of galaxies show alignment between their present-day AM and the orbital AM of their most massive (dominant) merger progenitor. This reorientation typically results in changes of around 50\% in the galaxies' specific AM, with the most significant shifts occurring in galaxies that were initially highly misaligned. We find only a weak influence from the second most massive merger, and negligible impacts from surviving satellites. We show that accreted stellar halos encode the history of this reorientation. Driven by the same accretion event, the main bodies of galaxies and their stellar halos tend to co-align, with 81$\pm2$\% of TNG-50 stellar halos showing prograde rotation relative to the galaxy. This signature will be detectable through major-axis kinematics with 30-meter class telescopes for Milky Way mass galaxies, offering a valuable observational test of this picture. While halo rotation directly constrains the specific AM of mergers within the last $\sim7$\,Gyr, this kinematic `memory' is largely erased for older and more radial events. Consequently, the Milky Way itself appears to be a notable exception to the general merger-driven trend: TNG-50 analogs with early, radial, and low angular momentum dominant mergers affect present-day disk orientation minimally. The current MW disk orientation may instead reflect the accumulated influences of gas accretion or dark matter torques.

\end{abstract}

\keywords{Galaxies (573) --- Galaxy stellar halos (598) --- Galaxy mergers (608) --- Galaxy evolution (594) }


\section{Introduction}

In the $\Lambda$CDM paradigm of galaxy formation, galaxies form through the cooling and infall of gas into dark matter halos \citep{Somerville2015}. In this process, enough angular momentum is preserved that disks form very commonly in galaxies with stellar masses $M_*>10^9 M_{\odot}$ \citep{Fall1980,MMW1998}. Consequently, the properties of galactic disks are important constraints for our picture of how halos acquire angular momentum, and the processes of gas cooling, star formation, and feedback \citep[e.g.,][]{Romanowsky2012,Fall2018,Posti2019}. Angular momentum evolution affects not only the magnitude of angular momentum but also its direction, which appears to lead to subtle patterns in the orientation of disks compared to their surroundings that reflect the ways in which galaxies gain their angular momentum \citep[e.g.,][]{Welker2014,Barsanti2022,Samuroff2023}.

Yet, galaxy merging --- another central aspect of galaxy formation in a $\Lambda$CDM Universe --- complicates this picture of disk growth. Merging affects many of the observable features of galaxies: they frequently temporarily enhance star formation \citep[e.g.,][]{Barnes1996,Ellison2008}, impact the formation of bulges and bars \citep[e.g.,][]{Mihos1996,Lang2014}, and crucially, may warp, thicken, or disrupt galactic disks \citep[e.g.,][]{Kaz09,Naab2014}. 

A clear but lesser-appreciated consequence of merging and satellite accretion is {\it disk tilting} \citep{HuangCarlberg1997,Kaz09,Welker2014,Dodge2023,Dillamore2022}. Disk tilting acts hand in hand with disk heating, in a way that depends on satellite orbit: disk tilting is a coherent response, particularly prominent at intermediate orbital inclinations, while the heating in the vertical direction is the dominant mode for e.g., polar orbits with little angular momentum in the disk plane \citep{Kaz09,Dodge2023}. \citet{Dillamore2022} uses the high resolution {\sc artemis} simulations --- a set of cosmological zoom-in simulations of $8\times10^{11} < M_{200}/M_{\odot} < 2\times10^{12}$ --- to show that re-orientation of galactic disks is expected to be common, with 12 out of their 15 galactic disks reorienting themselves to align with the angular momentum of the incoming satellite, shifting as rapidly as $\sim60^{\circ}$\,Gyr$^{-1}$. With the modest samples of simulated galaxies that we have studied to date, it appears that any picture of angular momentum evolution of galactic disks is incomplete without considering the impacts of galaxy mergers and accretions. 

Although disk tilting in response to satellite accretion and merging should be common, its effects will be extremely challenging to observe. Subtle preferential alignments with filaments and large scale structure have been measured \citep[e.g.,][]{Tempel2013,Barsanti2022,Samuroff2023}, but information about the orbit of the merging satellite is not directly available, but instead is encoded only in the orientation of the galaxy compared to its local environment \citep{Welker2014,Tempel2015,GV2019}. Galaxy warps could encode information about disk tilting, but can also result from misaligned gas accretion \citep{Earp2017,Earp2019} and halo figure rotation \citep{Johri_etal_2026}. In the Milky Way, other probes may exist, such as systematic trends in the properties of stellar streams as a function of location in the Galaxy \citep{Nibauer2024}.  

Stellar halos may offer novel insight into this avenue for angular momentum acquisition by galaxies. Stellar halos contain most of the debris from satellite accretions and mergers \citep[e.g.,][]{Bullock2001,BullockJohnston2005,Cooper2010}, and because this accreted material is deposited within the host's gravitational potential, stellar halos also naturally reflect the spin and shape of the innermost parts of the dark matter halo \citep{Bailin2005,Dillamore2022,Emami2021}. Simulations and observations indicate that most of the mass in a given stellar halo comes from the largest past merger event \citep[e.g.,][]{Cooper2010,Deason2016,Harmsen2017} --- a merger that we will term the dominant merger (following \citealt{DSouza2018_MNRAS}). It is reasonable to hypothesize that the kinematics of a stellar halo should reflect the orbital parameters of the merging satellite, including its angular momentum (see also \citealt{Gomez2017} and \citealt{Obreja2018}). 

This paper therefore has two goals. Firstly, we aim to explore galaxy reorientation via mergers and accretions using TNG-50 \citep{Pillepich2019}, a cosmological simulation that offers a diverse sample of $>$10$\times$ more simulated systems than have been previously studied, constraining how sensitive this phenomenon is to simulation specifics, galaxy morphology, and generalizing from earlier studies. Secondly, we quantify the kinematics of the accreted stellar halos in TNG-50, revealing that most stellar halos have modest angular momentum in the direction of overall galaxy rotation, which should be observable with kinematic surveys of stellar halos along their major axes. We show that the preferential alignment of halo and overall galaxy rotation reflects the reorientation of galaxies during and after their dominant merger events, offering a new and powerful test of mergers as a way of delivering angular momentum to galaxies.

\section{Quantifying the angular momentum, satellites and stellar halos of a TNG-50 simulated sample of Milky Way-mass galaxies}
\begin{figure*}[t]
    \centering
    \BeginAccSupp{method=escape,Alt={Two-panel schematic showing the reorientation of a galaxy and its stellar halo to align with a merger's orbital axis. Left panel shows initial misalignment at infall; right panel shows present-day alignment.}}    \includegraphics[width=\linewidth,interpolate=true]{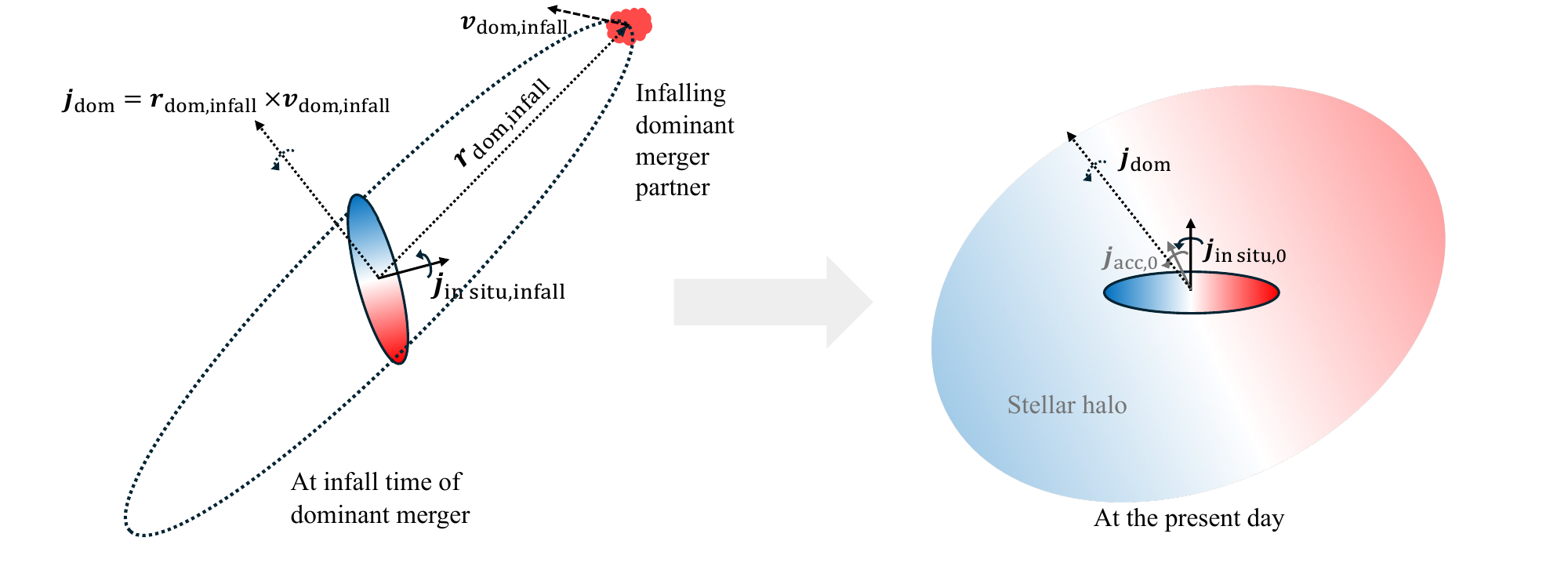}
    \EndAccSupp{} \vspace{-0.2cm} 
    \caption{A schematic illustration of the reorientation of a galaxy's AM by merging with its dominant merger partner. At the time of infall, the dominant merger partner is infalling with velocity $\bm{v}_{\mathrm{dom,infall}}$ (which is redshifted into the page, as signified by the color of the infalling satellite) with position $\bm{r}_{\mathrm{dom,infall}}$ and specific angular momentum $\vec j _{\mathrm{dom,infall}} = \vec r_{\mathrm{dom,infall}} \bm{\times} \vec v_{\mathrm{dom,infall}}$. As this merger proceeds and creates a stellar halo, the galaxy's angular momentum $\vec j _{\mathrm{in\,situ,infall}}$ reorients to $\vec j _{\mathrm{in\,situ,0}}$; the stellar halo also has $\vec j _{\mathrm{acc,0}}$ that in part reflects the angular momentum of the dominant merger.}
    \label{fig:diskreorient}     \vspace{0.05cm}
\end{figure*}

We use the TNG-50 simulation for this work \citep{Pillepich2019}. TNG-50 combines a substantial volume (51.7\,Mpc)$^3$ with  resolutions approaching those achieved by zoom-in simulations (baryonic mass resolution of $\sim 8.5 \times 10^4\,M_{\odot}$ and spatial resolutions $\sim$300\,pc). The simulation box contains hundreds of galaxies with similar stellar masses to the Milky Way \citep{Pillepich2024}, resolving even low-mass stellar haloes ($\sim 10^8$ to $10^9\,M_{\odot}$) with the necessary number statistics to measure spatially-resolved stellar halo kinematics. We select galaxies that broadly encompass the stellar mass of the Milky Way, choosing stellar masses\footnote{We adopt stellar masses within a 30\,kpc aperture as a reasonable estimate of the observable, see also \protect\citet{Pillepich2019} and \protect\citet{Pillepich2024}.} between $10^{10}M_\odot \le M_{\ast, \text{30\,kpc aperture}} \le 2\times10^{11}M_\odot$. We choose halo dark matter (DM) masses ($M_{halo}$) $M_{200c}\le10^{13}\,M_\odot$, and choose central galaxies by ensuring that the subhalo's dark matter halo mass is within 0.2\,dex of the group's dark matter halo mass. This results in a sample of 492 simulated systems.  

For the purposes of this work, we choose to study the kinematics of the main bodies of galaxies using {\it in situ} star particles, and their halos using accreted star particles. We tag individual particles as accreted or {\it in situ} following \citet{DSouza2018_MNRAS}. In brief, particles formed in the main progenitor branch of the galaxy (using the SUBLINK merger trees; \citealt{Rodriguez-Gomez2015}) are deemed to be {\it in situ} stars. The other stellar particles found in the subhalo of the main galaxy are tagged as ‘accreted’; this definition is consistent with \citet{Rodriguez-Gomez2016}.  

Previous studies have demonstrated that most of the mass in a stellar halo is delivered by its most massive previous merger partner \citep{Cooper2010,Deason2016,Harmsen2017,DSouza2018_MNRAS,Monachesi_2019}. 
Consequently, as a first step, we identify each system's most massive completed merger partner, recording its infall time into the galaxy $t_{\mathrm{infall}}$, when it passes the virial radius of the main galaxy, and its merger time $t_{\mathrm{merger}}$, when it is no longer recovered as a galaxy. We record its specific angular momentum at the time of infall, $\vec j _{\mathrm{dom,infall}} = \vec r_{\mathrm{dom,infall}} \bm{\times} \vec v_{\mathrm{dom,infall}}$, where $\vec r_{\mathrm{dom,infall}}$ is the position of the infalling satellite relative to the main galaxy at the time of infall, and $\vec v_{\mathrm{dom,infall}}$ is their relative velocity (see Fig.\ \ref{fig:diskreorient}). 

In order to understand the effect of this merger on the main galaxy, we record the specific angular momentum vector of the stellar particles in the main galaxy:
\begin{equation}
    \vec j _{*} = \frac{\sum_i m_i \vec r_i {\bm{\times}} \vec v_i}{\sum_i m_i}, 
\end{equation}
where the positions and velocities are measured relative to the center of mass of the main galaxy, and we sum over all star particles in the main galaxy. This can be calculated at any time in the main galaxy's evolutionary history, but we will focus on its angular momentum at the present day $\vec j _{*,0}$ and at the time of infall $\vec j _{\mathrm{*,infall}}$. To give a rough sense of the expected order of magnitude for $|\vec j|$, we note that $|\vec j| \sim 2 r_h v_c$ \citep{Romanowsky2012}, where $r_h$ is the exponential scale length of a galaxy and $v_c$ is its circular velocity; for the Milky Way (with $r_h \sim 2.6$\,kpc; \citealt{Licquia2016} and $v_c \sim 230$\,km\,s$^{-1}$; \citealt{Eilers2019}) this value is roughly $|\vec j_{\mathrm{MW}}| \sim 1200$\,kpc\,km\,s$^{-1}$. 

In order to explore differences between the evolution of the AM of galaxies as a function of morphology, we classify systems roughly into `disks' and `non-disks', using the fraction of stars in high angular momentum orbits (see e.g., \citealt{Snyder2015} and \citealt{Rodriguez-Gomez2017} for more discussion). A `circularity' parameter $\epsilon$ is calculated for each stellar particle $i$ where $\epsilon_i = j_{z,i}/j_{z,max}(E_i)$, where $j_{z,i}$ is the specific angular momentum around the minor axis of the galaxy (defined using the galaxy's AM vector), and $j_{z,max}(E_i)$ is the maximum angular momentum of the 100 star particles with the most similar energy to the star particle in question $E_i$. We define `disks' to be those systems with a fraction of stars with $\epsilon>0.7$ equal to or larger than the median value (0.48, so just under half the stars are in rather circular orbits), and `non-disks' less than that median value. The adoption of this particular definition is not critical; other classification measures (e.g., just splitting by specific angular momentum) yield very similar results. 

Later, when we explore the kinematics of the accreted stellar halo of a galaxy as a possible observational probe of satellite angular momentum, we calculate both the angular momentum of the accreted stars and an example mock measurement that could be made for stellar halos. We define a measure of the typical angular momentum of accreted particles within ranges of galactocentric radius:
\begin{equation}
    \vec j _{\mathrm{acc}} = \frac{\sum_i m_i \vec r_i \bm{\times} \vec v_i}{\sum_i m_i}, 
\end{equation}
where we sum all the accreted star particles in the main galaxy with galactocentric radii in a given radial range, typically within 40\,kpc. 

Motivated by our interest in stellar halo angular momentum as an observational probe of disk realignment by satellite accretion, we choose to sidestep the complexity of accreted star velocity fields and focus on the kinematics of stellar halo particles along the galaxy's projected major axis. These mock measurements are appropriately affected by kinematic substructure, but these substructures are not our focus in this paper. For our mock halo kinematic measurement, we measure the median velocity $v_{\mathrm{35,acc}}$ and velocity dispersion $\sigma_{\mathrm{35,acc}}$ (specifically, half of the difference between the 16th and 84th percentiles of the accreted stars in this region) of the accreted stars of a field 35\,kpc along one of the galaxy's major axes, with a field size of $(10\times10)$\,kpc$^2$. We choose to adopt the convention that the velocities are positive if they are in the same direction as the rotation of the {\it in situ} stars on that side of the galaxy (i.e., their orbits are prograde). This roughly mirrors the kind of measurements of halo kinematics that might be available for external galaxies in the future with the advent of 30-meter class telescopes, as foreshadowed by the deep stellar halo kinematics of a 1\,kpc$^2$ field 35\,kpc along the major axis of NGC 4945 from \citet{Beltrand2024}, and a deep stellar halo kinematic field along the major axis of NGC 253 probing to projected major axis radii $r_{\mathrm{major}}<20$\,kpc (K. Gozman et al., in preparation). We note that none of the conclusions of this paper are materially affected by the detailed choice of the position along the major axis at which we measure the stellar halo kinematics, or indeed which side of the galaxy we choose to explore. 

We choose to discard those galaxies that have such small dominant mergers that they have peak dark matter masses $<1$\% of the dark matter mass of the central galaxy, or insufficient particles to measure stellar halo kinematics, omitting 21 systems, for a total sample of 471 systems.

\section{Disk reorientation during and after massive satellite accretion in TNG-50}

\begin{figure*}[t]
    \centering
    \BeginAccSupp{method=escape,Alt={Two histograms. Left: specific angular momentum shifts from a zero-centered peak at infall to a positive peak today. Right: galaxy alignment moves from a uniform distribution towards a sharp peak at cosine theta equals 1 for the dominant merger.}}
    \includegraphics[width=\linewidth]{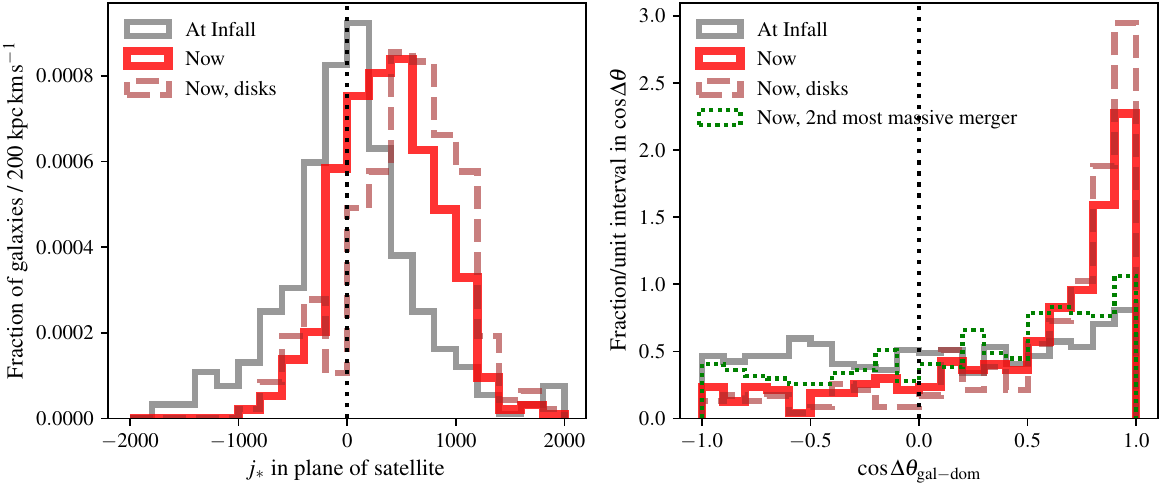} 
    \EndAccSupp{}  \vspace{-0.2cm}  
    \caption{{\it Left:} The distribution of the component of the galaxy specific angular momentum in the direction of the angular momentum of the infall of the dominant progenitor; the $y$-axis is the fraction of galaxies in a single bin of specific AM. There is little alignment between the galaxy's angular momentum and the infalling satellite at infall time (gray). In contrast, the present-day galaxy AM shows more alignment (red), with 80\% of systems having positive specific AM relative to the satellite infall plane. Dark red shows the present-day alignment of the subset of disk galaxies (as defined in the text). {\it Right:} The distribution of $\cos \Delta\theta_{\mathrm{gal-dom}}$, the cosine of the angle between the {\it in situ} stars' AM and dominant merger orbital AM, showing a clear preference towards alignment at the present day. The distribution of $\cos \Delta\theta_{\mathrm{gal-2nd}}$, the angle between the galaxy AM and that of the second most massive merger partner, is shown in a thin green dotted line. 
    The AM content of present-day galaxies was affected by mergers, particularly with the dominant merger partner. }
    \label{fig:disk_kin}     \vspace{0.05cm}
\end{figure*}

\subsection{Disk angular momentum is correlated with the orbital angular momentum of the dominant merger progenitor}

We start by showing that at the present day the angular momentum of galaxies is more aligned with the orbital angular momentum of the most massive merger progenitor than chance alone allows; this is true for all galaxies, and is more pronounced for those galaxies with the largest degree of rotational support. The left-hand panel of Fig.\ \ref{fig:disk_kin} shows the component of the {\it in situ} stars' specific AM in the direction of the dominant merger's orbital AM, $j_{\mathrm{gal-dom}} = \vec{j}_{\mathrm{*,galaxy}} \cdot \hat{\vec \jmath }_{\mathrm{dom}}$, where $\hat{\vec \jmath }_{\mathrm{dom}}$ is the unit vector in the direction of the dominant merger partner's orbital AM. The AM of the {\it in situ} component of the main galaxy at the time of infall is poorly correlated with the orbital AM of the dominant merger partner (gray). In contrast, {\it in situ} stars in present-day galaxies tend towards closer alignment with the dominant merger partner's orbital AM (red), with 80\% of galaxies (377/471) having positive specific AM. Disks align with dominant mergers more effectively still (dashed brown lines), with 86\% (202/234) of disks having positive specific AM in the direction of dominant merger orbital AM. 

The relative angle between the {\it in situ} stars' AM and dominant merger's infall orbital AM is shown in the right-hand panel of Fig.\ \ref{fig:disk_kin}. A random distribution in angles should be uniform in $\cos \Delta \theta_{\mathrm{gal-dom}}$; this is close to true at the infall time (gray). The present-day galaxy orientation shows more correlation with dominant merger orbital infall AM, both for all galaxies (red) and disks (dashed brown lines).  It is worth emphasizing that there is generally still significant misalignment between the orientation of the {\it in situ} stars' AM and the dominant merger's AM. The median value of $\cos \Delta \theta_{\mathrm{gal-dom}}$ at the present day for all galaxies is 0.69$^{+0.01}_{-0.03}$; i.e., the median misalignment is 46$\degr$. Strong alignment is not that common; only 27\% of systems have $<30\degr$ misalignment. This emphasizes that the contribution of satellite merging to galactic AM evolution is significant but is just one of several important factors that influence a galaxy's AM. 

\begin{figure*}[t]
    \centering
    \BeginAccSupp{method=escape,Alt={Six-panel scatter plot showing the change in specific angular momentum. Panel d shows a strong linear correlation with orbital inclination, and panel a shows a strong linear correlation with the magnitude of the final angular momentum. Panels b, c, e, and f show scatter around a mean positive change in angular momentum for mass, time, orbital angular momentum, and circularity.}}
        \includegraphics[width=\linewidth]{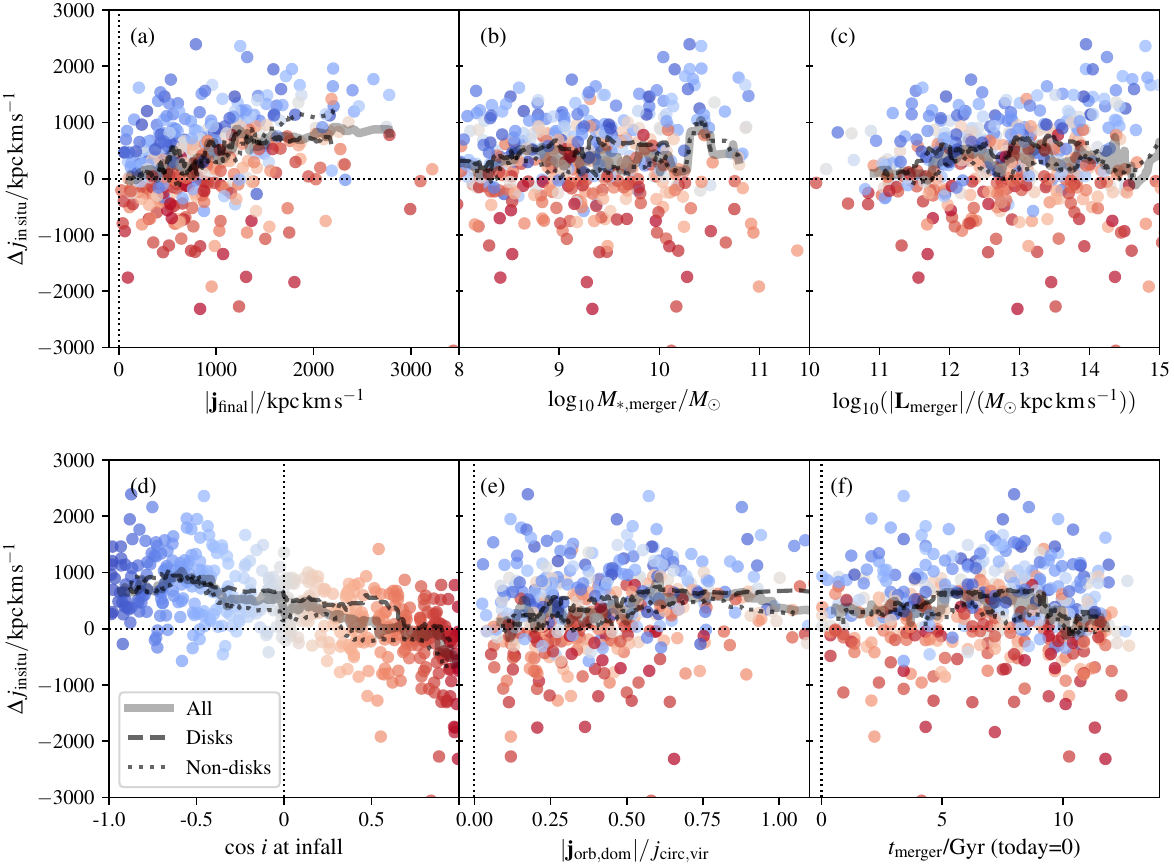}
    \EndAccSupp{} \vspace{-0.3cm}
    \caption{The change in the main galaxy's specific angular momentum as a function of several parameters: {\it a)} the amplitude of the final specific angular momentum, {\it b)} the stellar mass of the dominant merger, {\it c)} the angular momentum of the dominant merger, {\it d)} the inclination of the orbit compared to the infall orbit plane (parameterized as $\cos i$), which is the color code in all panels, {\it e)} the eccentricity of the satellite's initial orbit at the time of infall, and {\it f)} the merger time (where $t_{\mathrm{merger}}=0$ is the present day). Running medians of the nearest 25 points are shown in the thick gray line for all galaxies, and in the thinner dashed/dotted lines for disks and non-disks respectively. }
    \label{fig:delta_AM}     \vspace{0.05cm}
\end{figure*}

We note that these results are largely unaffected by choices of which stars in the main bodies of galaxies we choose for analysis. In particular, we note that old {\it in situ} stars, formed before the infall of the dominant merger, also reorient; 80\% of the old stars have positive AM in the direction of the dominant merger's orbital AM. The old {\it in situ} stars are generally well aligned with the younger {\it in situ} stars formed after the dominant merger's infall time, with 90\% of them being aligned to within 30\degr and 78\% of them being aligned to within 10\degr. This emphasizes that the processes that reorient galaxies are dynamical in origin, affecting both old and young stars.

\subsection{Factors influencing the degree of reorientation}

Are there obvious trends in the degree of reorientation as a function of merger parameters? In Fig.\ \ref{fig:delta_AM}, we show the change in the specific angular momentum of the {\it in situ} stars in the direction of the dominant merger's angular momentum vector, as a function of various parameters. Running medians of the nearest 25 points in the $x$-parameter are shown in thick gray lines; the running medians of disks and non-disks are shown in thin dashed and dotted lines, respectively. There are quantitative differences in the reorientations experienced by disks and non-disks, while the qualitative behaviors of the two classes of galaxy are essentially the same. This underlines that galaxy reorientation by mergers is a generic phenomenon that does not depend critically on the existence of a prominent disk component. 

We summarize the distribution of specific AM at infall and the present day, along with the resulting changes for different morphological subsets, in Table \ref{tab:am_percentiles}.  Although there is a lot of scatter, the typical main galaxy shows a change in specific angular momentum of around $\sim$1/2 of both its initial and final angular momentum. While disks experience $\sim200\,\mathrm{kpc\,km\,s}^{-1}$ more change in specific AM than non-disks, both types of galaxy experience reorientation in response to their dominant merger. 

\begin{table}
\centering
\caption{Specific Angular Momenta and Merger-Induced Changes} 
\label{tab:am_percentiles} %\vspace{-0.2cm}
\BeginAccSupp{Alt={Table of percentiles of specific angular momentum. Medians for all galaxies are 720 at infall and 690 at redshift zero. The median change in angular momentum is 330 for all galaxies, but higher for disks at 510.}}
\begin{tabular*}{\columnwidth}{@{\extracolsep{\fill}}lccc@{}}
\toprule
Component / Subset & 25th & 50th (Median) & 75th \\ 
\midrule
\multicolumn{4}{@{}l}{\textbf{Absolute Specific AM ($j_{\mathrm{in{\,}situ}}$)}} \\
\midrule
\quad All Galaxies at $t_{\mathrm{infall}}$ & 380 & 720 & 1260 \\
\quad All Galaxies at $z=0$                 & 430 & 690 & 980 \\
\quad Disks at $z=0$                        & 650 & 880 & 1100 \\
\quad Non-disks at $z=0$                    & 290 & 480 & 730 \\
\midrule
\multicolumn{4}{@{}l}{\textbf{Change in Specific AM ($\Delta j_{\mathrm{in{\,}situ}}$)}} \\
\midrule
\quad All Galaxies                          & $-110$ & 330 & 780 \\
\quad Disks                                 & 30     & 510 & 810 \\
\quad Non-disks                             & $-200$ & 232 & 660 \\
\bottomrule
\end{tabular*}

\vspace{0.15cm}
\justifying

    \footnotesize 
    \textit{Notes:} --- Percentiles of the magnitude of the specific angular momentum ($j_{\mathrm{in{\,}situ}}$; top rows) and the change in angular momentum between the time of satellite infall and the present day ($\Delta j_{\mathrm{in{\,}situ}}$; bottom rows) in the direction of satellite orbital AM. All values are in units of $\mathrm{kpc{\,}km{\,}s}^{-1}$. 
\EndAccSupp{}
\end{table}

There is a strong correlation between the change in specific angular momentum in the direction of the infalling satellite orbit and the magnitude of the final {\it in situ} specific angular momentum (panel a; Fig.\ \ref{fig:delta_AM}). There is no correlation between the change in the main galaxy's specific angular momentum and the magnitude of its initial angular momentum (not shown). TNG-50 therefore predicts that those systems with large specific angular momentum today are dominated by those that had large, satellite-induced changes in their angular momentum. Put differently, mergers appear to increase the angular momentum of many galaxies, and this effect should be particularly prominent for galaxies with the highest specific angular momentum today.  

The next panels explore satellite/merger parameters that one may expect to influence the degree of disk reorientation. Notably, many of these parameters appear to have little coherent effect on the degree of reorientation. 

The inclination of the orbit compared to the infall orbit plane (parameterized as $\cos i$, shown as the color code in all panels of Fig.\ \ref{fig:delta_AM}) shows the strongest correlation with the degree of reorientation, in an intuitive sense --- initially more misaligned systems show a larger degree of reorientation (panel d, Fig.\ \ref{fig:delta_AM}). Perhaps amusingly, those galaxies that were aligned at infall time with $\cos i \gtrsim0.9$ typically drift in their orientation away from the orbital AM of the infalling dominant merger, underscoring that mergers with satellites are only one of many factors that set galaxy orientation, and that when the influence of the satellite is unimportant (when it is aligned anyway), other factors dominate the evolution of a galaxy's orientation. This trend holds for both disks and non-disks, although disks reorient more than non-disks. 

Other parameters show little apparent effect. There is a weak tendency for satellites with lower orbital eccentricity at the infall time (parameterized as the angular momentum $|\vec j_{\mathrm{orb,dom}}|/j_{\mathrm{circ,vir}}$, the fraction of the angular momentum that a satellite has compared to the magnitude of the specific angular momentum expected for a satellite in a circular orbit at that radius $j_{\mathrm{circ,vir}}$) to impart more angular momentum (panel e). Mergers with little AM impart little AM (panel c). The earliest mergers $\gtrsim 10$\,Gyr ago have little effect on today's galaxy orientation (panel f). And, perhaps most surprisingly, the stellar mass of the dominant merger correlates very little with the degree of reorientation (panel b). This last effect may be made easier to understand by noting that the total (stellar$+$dark matter) mass of the dominant mergers is $\gtrsim10^{10} M_{\odot}$, rivaling the mass of the main bodies of galaxies --- these infalling satellites dominate the total AM budget of the system (as also discussed by \citealt{Welker2014} and \citealt{Dillamore2022}). 

Although $\cos i$ is clearly the most prominent correlate, other parameters could correlate with the magnitude of the degree of reorientation $\Delta j_{\mathrm{in\,situ}}$. We explored this issue in a few ways. Performing multi-variate linear regression allowing all five parameters ($\cos i$, $|\vec j_{\mathrm{orb,dom}}|/j_{\mathrm{circ,vir}}$, stellar mass, angular momentum, and merger time) showed a strong correlation with $\cos i$ and correlations with $<2\sigma$ significance for all other parameters. We attempt to identify the most significant predictors of $\Delta j_{\mathrm{in\,situ}}$ by using the Least Absolute Shrinkage and Selection Operator (LASSO; \citealt{tibshirani96regression}, see also \citealt{acquaviva2023}). In contrast to standard linear regression, LASSO imposes a penalty on the regression coefficients, allowing for automated feature selection by shrinking the coefficients of less impactful variables to zero. We scaled all variables to have a mean of zero and a standard deviation of 1 for the fit; for reporting coefficients we performed the inverse scaling. The tuning parameter $\lambda$ was optimized to be $\lambda \sim 100$ using 5-fold cross validation. Although only two parameters returned non-zero coefficients ($\cos i$ and $|\vec j_{\mathrm{orb,dom}}|/j_{\mathrm{circ,vir}}$), fitting of bootstrap samples gave $<1.5\sigma$ confidence in the slope of $|\vec j_{\mathrm{orb,dom}}|/j_{\mathrm{circ,vir}}$. Therefore, to excellent approximation, $\Delta j_{\mathrm{in\,situ}}$ is thought of best as a function of $\cos i$ alone:
\begin{equation}
\Delta j_{\mathrm{in\,situ}} = 26\pm80 - (610\pm50) \cos i, 
\end{equation}
with a scatter of $\sigma = 620\,\mathrm{kpc\,km\,s^{-1}}$ around this fit. The correlation with $\cos i$ only accounts for 35\% of the variance in $\Delta j_{\mathrm{in\,situ}}$, the other 65\% is due to other factors. We note that these findings hold for disks and non-disks separately --- the degree of reorientation appears to vary systematically with $\cos i$, with most of the variance contributed by factors that do not appear in our analysis. We conclude that the inclination of the infalling dominant merger is the most important factor affecting the degree of galaxy reorientation from merging, but with a great degree of variation that underlines the idea that many factors affect galaxy orientation. 

\subsection{The influence of other mergers or satellite accretions}
\label{sec:others}

It is interesting to ask whether other mergers or satellite accretions have an impact on galaxy orientation. In brief, we find that the second most massive merger correlates weakly with galaxy orientation, while correlations with surviving satellites are negligible. 

We must first check whether the orbital angular momenta of other satellites are correlated with the orbital angular momentum of the dominant merger. If they were correlated, then any correlation of {\it in situ} stars' AM orientation with the other satellites or mergers could be attributed to the dominant merger. We find that the orbital AM of the second most massive merger, and those of the two most massive surviving satellites, are uncorrelated with the orbital AM of the dominant merger. This implies that any correlations between galaxy orientation and satellite AM reflect {\it only} the impact of that satellite, not the dominant merger partner. 

The {\it in situ} stars' AM is weakly correlated with the orbital AM of the second most massive merger (Fig.\ \ref{fig:disk_kin}, thin green line). The median $\cos \Delta \theta_{\mathrm{gal-2nd}} = 0.33\pm0.05$, considerably smaller than the degree of alignment with the dominant merger $\cos \Delta \theta_{\mathrm{gal-dom}} = 0.69^{+0.01}_{-0.03}$. The degree of alignment between the {\it in situ} AM and the orbital AM (at infall) of surviving satellites is $\cos \Delta \theta_{\mathrm{in\,situ-dom}} \lesssim 0.1$ --- small effects, which are consistent with being drawn from a uniform distribution $\sim5$\% of the time. 

This finding (counter-intuitively?) clarifies the role of merging in impacting the AM of galaxies. Completed mergers contribute to the evolution of the AM of galaxies, where the most massive merger tends to have the most impact --- this holds for dominant mergers with a wide range of stellar masses (panel b) of Fig.\ \ref{fig:delta_AM}). The second most massive merger also has an effect, but because it has (by definition) lower mass, it has a smaller impact than the dominant merger. Interactions with surviving satellites do not appear to significantly affect the orientation of galactic AM.

\section{Stellar halo kinematics as a (complex) diagnostic of dominant merger orbit and reorientation}

\begin{figure*}[t]
    \centering
    \BeginAccSupp{method=escape,Alt={Two-panel histograms of cosine of the angle between angular-momentum vectors. Left: dominant merger orbital angular momentum vs in-situ stars, stellar halo (several radii), and dark matter within 50 kpc—most are strongly aligned; the surviving satellite is not. Right: in-situ stellar angular momentum vs the same components—strong alignment for in-situ, halo, and dark matter; weak for the surviving satellite.}}%
    \includegraphics[width=\linewidth]{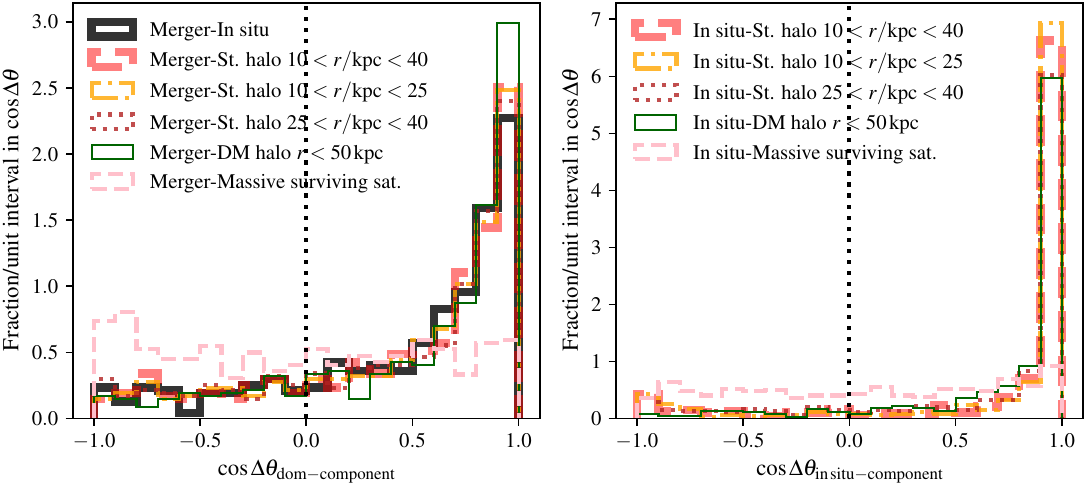}%
    \EndAccSupp{} \vspace{-0.2cm} 
    \caption{{\it Left:} The distribution of $\cos \Delta\theta_{\mathrm{dom-component}}$, the cosine of the angle between the dominant merger orbital AM and various components: the {\it in situ} stars (black solid line, same as red in Fig.\ \ref{fig:disk_kin}), the accreted stellar halo in various radial ranges (shades of orange/red, with various linestyles), dark matter $r<50$\,kpc (thin solid green), and the most massive surviving satellite (thin pink dashed). In almost all cases, these components show a similar degree of preference towards alignment with the dominant merger AM at the present day; the exception is the most massive surviving satellite which shows essentially no correlation with the dominant merger. {\it Right:} The distribution of $\cos \Delta\theta_{\mathrm{in\,situ-component}}$, the cosine of the angle between the in situ stars AM and various components. The alignment in AM content between the {\it in situ} stars, accreted stars and DM is usually very strong; the correlation with the AM of the most massive surviving satellite is very weak.  }
    \label{fig:dir_AM}     \vspace{0.05cm}
\end{figure*}

Galaxy reorientation through mergers is an important avenue of AM acquisition, with changes of $\sim1/2$ of a galaxy's specific AM taking place through the merger process. It is valuable to ask what observable components of galaxies or halos may reflect this. It is most natural to focus on the two components that are in great part or entirely accreted in origin --- e.g., surviving satellites and stellar halos (the debris from merged satellites). 

The left-hand panel of Fig.\ \ref{fig:dir_AM} shows the distribution of the dot product $\hat{\vec \jmath}_{\mathrm{dom}}\cdot\hat{\vec \jmath}_{\mathrm{component}}$, for the {\it in situ} stars (black, see also Fig.\ \ref{fig:disk_kin}), the stellar halo in different radial ranges ($10<r/\mathrm{kpc}<40$ in thick red; inner halo $10<r/\mathrm{kpc}<25$ in orange thinner dash-dotted, and outer halo $25<r/\mathrm{kpc}<40$ in thin dotted dark red), dark matter ($r<50\,\mathrm{kpc}$, thin solid green line), and the infall AM of the most massive surviving satellite (dashed thin pink line). This is equivalent to the cosine of the relative angle between the angular momenta of the dominant merger and a galaxy component. Unlike the most massive surviving satellite, one can see that {\it in situ} stars, accreted stars, and dark matter all show similar degrees of alignment between their AM and the orbital AM of the dominant merger. 

The right-hand panel of Fig. \ref{fig:dir_AM} shows that the {\it in situ} stars, stellar halo, and inner dark matter (DM) halo exhibit strong mutual alignment at the present day. This coherence suggests a global dynamical reorientation during massive accretion events, where the stellar halo effectively reflects the shape and kinematics of the underlying dark matter distribution \citep{Bailin2005, Earp2019, Emami2021}. Notably, these components align more closely with one another than any individual component does with the initial orbital AM of the dominant merger. While the halo `remembers' the merger, that memory is filtered through the subsequent dynamical relaxation of the entire system. 

\begin{figure}[t]
    \centering
    \BeginAccSupp{method=escape,Alt={Plot of specific angular momentum for in situ and stellar halo components versus the dominant merger. Both components exhibit a positive trend that closely tracks a reference line where the component angular momentum is one-sixth that of the dominant merger.}}
    \includegraphics[width=\columnwidth]{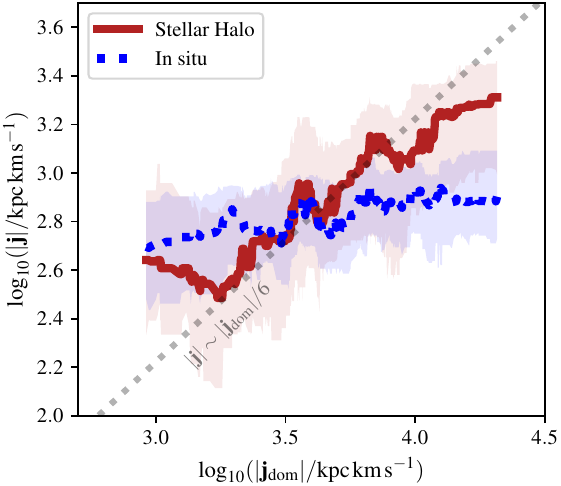}
    \EndAccSupp{}     \vspace{-0.2cm} 
    \caption{The running median (line) and interquartile ranges (shaded regions; from nearest 51 neighbors) of the magnitude of the specific AM for accreted stellar halo particles (red solid line) and {\it in situ} stars (dotted blue line), as a function of the magnitude of the specific orbital AM of the dominant merger. A schematic line showing the trend expected if the stellar halo specific AM is $\sim 1/6$ of the dominant merger's AM is shown with the dotted line.   }
    \label{fig:amp_AM}     \vspace{0.05cm}
\end{figure}

Turning to the {\it magnitude} of the stellar halo specific AM, we find a broad correlation with dominant merger orbital specific AM (Fig.\ \ref{fig:amp_AM}, solid red line with interquartile range of the distribution). Unlike the {\it in situ} star specific AM (dotted blue line with interquartile range of the distribution), halo AM looks to be correlated almost linearly with dominant merger AM, where $\log_{10}{|\vec j_{halo}|} \sim \log_{10} |\vec j_{dom}| - 0.8^{+0.4}_{-0.5}$ (median and 16/84 percentiles), albeit with roughly 0.5\,dex (a factor of three) scatter. Thus, our results suggest that the direction of the stellar halo AM contains information about direction of the infalling satellite orbital AM, subject to global responses that bring it into closer alignment with the {\it in situ} stellar distribution. The stellar halo AM magnitude remains an informative, though noisy, probe of the scale and impact parameters of the dominant merger event.

\section{Predictions for stellar halo observations}

We have found that stellar halos, in principle, encode information about the global dynamical response of a galaxy to its dominant merger and the orbital angular momentum of that merger. We now briefly examine mock halo kinematic observations to explore possible observational signatures of dominant merger angular momentum and galaxy reorientation. Many different measurements of halo kinematics could be suggested and no doubt will one day be carried out. For now, recognizing the extreme difficulty of stellar halo kinematic observations for external MW-mass galaxies that require them to focus on relatively small areas and samples (e.g., \citealt{Beltrand2024}, K.\ Gozman et al., in prep.), we will explore mock observations focused on particular halo regions --- generally only one field. 

We start by recalling important results from Fig.\ \ref{fig:dir_AM}. The stellar halo AM and {\it in situ} star AM are typically aligned well with each other, with the median $\cos \Delta_{\mathrm{in\,situ-halo}} = 0.978$  ($\Delta_{\mathrm{in\,situ-halo}} \sim 12$\degr), and 68\% of halos aligning to within 30{\degr} of the {\it in situ} star AM vector. This suggests that measurements relative to the disk plane will often be helpful. For this brief examination, we then choose to focus on galaxies in their edge-on projection. Furthermore, recognizing the observational expense and limited field of view of multiplexed spectrographs, we will measure stellar halo kinematics along only one side of the galaxy's projected major axis, where one would be most sensitive to the halo's likely AM content. 

\begin{figure}[t]
    \centering
    \BeginAccSupp{method=escape,Alt={Scatter plot showing a strong positive correlation between major axis mean halo velocity (x-axis) and stellar halo angular momentum (y-axis). }}
    \includegraphics[width=\columnwidth]{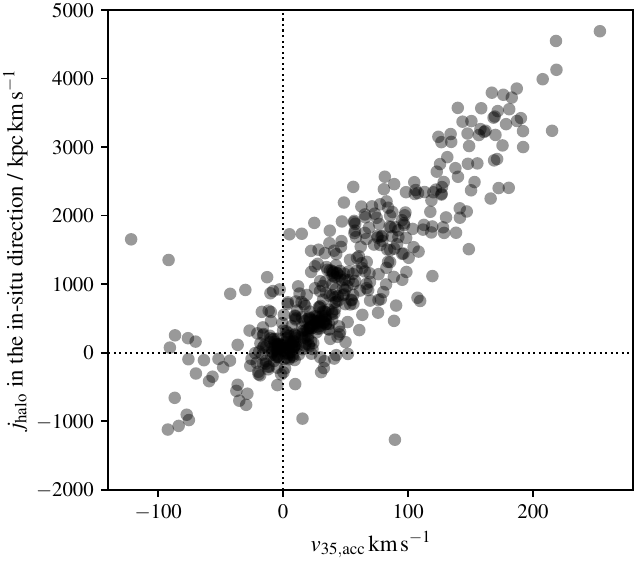}
    \EndAccSupp{}    \vspace{-0.2cm} 
    \caption{Major axis mean halo velocity ($x$-axis) and the stellar halo AM in the direction of the {\it in situ} AM ($y$-axis), for galaxies viewed from the edge-on perspective; positive velocities indicate prograde motion.   Major-axis velocity measurements (even along only one side of the galaxy) provide a useful measure of the stellar halo AM in the direction of {\it in situ} star rotation. }
    \label{fig:halo_vel_AM}     \vspace{0.05cm}
\end{figure}

We show an example of such a measurement for TNG-50 galaxies projected edge-on, where we show the projected mean velocity of halo stars in a $10\times10$\,kpc$^2$ region along one of the projected major axes (at a major axis distance of 35\,kpc, $v_{35,\mathrm{acc}}$) in relation to the stellar halo AM in the direction of the {\it in situ} star AM (Fig.\ \ref{fig:halo_vel_AM}. Although there is scatter between $v_{35,\mathrm{acc}}$ and stellar halo AM, even measurements along a single major axis carry important information about the direction and magnitude of stellar halo AM in the direction of the {\it in situ} star AM. Single-field halo kinematic measurements show clear signatures of halo AM and rotation --- measurements are predicted to tend towards being prograde, with the velocity offset from systemic velocity correlating with the magnitude of halo AM, without a great deal of correlation with other parameters. Recalling the predicted tendencies towards alignment between disks and stellar halos with each other, and between those components and the dominant merger (Figs.\ \ref{fig:dir_AM} and \ref{fig:amp_AM}), both the offset towards prograde motions and the magnitude of halo rotation show unique promise as probes of galaxy reorientation and AM acquisition from dominant mergers. 

\begin{figure}[t]
    \centering
    \BeginAccSupp{method=escape,Alt={Scatter plot comparing stellar halo velocity on the x-axis to the dominant merger's specific angular momentum on the y-axis, grouped into early and recent mergers. Running medians show a positive correlation that is stronger for recent mergers. Marginal histograms show the velocity distributions.}}
    \includegraphics[width=\columnwidth]{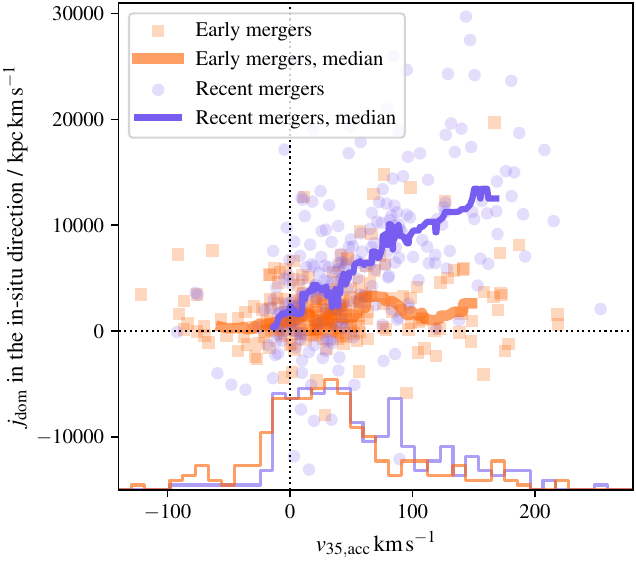}
    \EndAccSupp{}    \vspace{-0.2cm}     \caption{The relationship between stellar halo kinematics $v_{35,\mathrm{acc}}$ and the component of the dominant merger specific AM in the direction of the {\it in situ} star AM, split by dominant merger time (purple --- recent, $t_{\mathrm{ merger}}\le7.2$\,Gyr; orange --- early, $t_{\mathrm{ merger}}>7.2$\,Gyr). Running medians of the nearest 25 points in $v_{35,\mathrm{acc}}$ for each subset are shown as solid lines. We include a marginal histogram of the $v_{35,\mathrm{acc}}$ distribution of each population. While the scatter is substantial, halo kinematics correlates with dominant merger specific AM, particularly strongly for systems dominated by more recent mergers, offering an observational probe into the role of mergers in influencing the kinematics of galaxies.}
    \label{fig:halo_kin}
    \vspace{0.05cm}
\end{figure}

We illustrate this connection in Fig.\ \ref{fig:halo_kin}, where we correlate $v_{35,\mathrm{acc}}$ with the component of the dominant merger's AM in the direction of the {\it in situ} star AM --- this plot connects stellar halo observables with the orbital parameters of the dominant merger. There is a clear tendency towards larger dominant merger AM in the {\it in situ} star AM direction for higher $v_{35,\mathrm{acc}}$. Stellar halos are expected, broadly, to exhibit rotation in the same direction as the disk, with 25, 50 and 75 percentiles of $v_{35,\mathrm{acc}} = 8, 37,$ and 86\,km/s. Only 19\% of halos have $v_{35,\mathrm{acc}}<0$, rotating in the opposite sense to the disk. This preference towards preferentially prograde halo motions is a clear and testable observational consequence of disk reorientation by mergers. 

The time of the merger is an important second parameter; accordingly, we split the dataset by dominant merger time. 
For mergers more recent than the median merger time ($t_{\mathrm{merger}}\le t_{\mathrm{merger,median}}=7.2$\,Gyr; in purple, circles with the running median of the nearest 25 systems shown as a thin solid line), there is a strong preference towards prograde motions (85\% of younger halos are prograde). In addition, there is a clear but scattered correlation between the observable halo velocity $v_{35{\mathrm{,acc}}}$ and the component of the initial orbital angular momentum of the merging satellite, in the direction of {\it in situ} star rotation. Since merger time correlates with the age of stars in a stellar halo \citep{Harmsen2023}, and this can be constrained by an analysis of resolved stellar populations in halos \citep{DSouza2018_M31,Wang2020,Harmsen2023}, this implies that halo kinematics for these younger halos can offer a powerful observational probe of the orbital properties of the merger progenitor.  

The earlier mergers ($t_{\mathrm{merger}}>7.2$\,Gyr; in orange, squares and thick translucent line) paint a more complex picture. These halos are also predominantly prograde (74\% of older halos are prograde) --- a clear signature of disk reorientation by merging. Yet, there is very little correlation between halo rotation velocity and the dominant merger's initial orbital angular momentum, underlining the complexity of stellar halo formation and evolution. Accordingly, halo kinematic measurements of older halos will be less straightforwardly interpretable in terms of the dominant merger's orbital parameters, requiring much richer kinematic data to place meaningful constraints on merger orbit (as exists for the Milky Way; \citealt{Helmi_2018,Belukorov_2018}).  

\section{Discussion}

While many factors shape galactic orientation, we have shown that the main bodies of galaxies in the TNG-50 simulation undergo significant reorientation in response to massive mergers. Larger re-orientations are more likely for systems with larger initial differences in the initial galaxy AM and the orbital AM of the dominant mergers. 

Despite the fact that the magnitude of this effect is substantial, there are very few observable diagnostics that allow access to this phenomenon. The most useful probe of this process is the kinematics of stellar halos. Stellar halos also tend to reflect the orbital AM of the dominant merger, although the stellar halo and {\it in situ} AM alignments with each other are closer than either component has with the dominant merger AM as they are coupled to each other dynamically. The magnitude of the stellar halo AM correlates with a factor of 2.5--3 scatter with the magnitude of the dominant merger's orbital AM. Appreciating that many choices would be possible, we focused on a simple stellar halo kinematic measure of the median stellar halo velocity along one of the galaxy's minor axes, finding that this median velocity correlates well with the component of the stellar halo AM in the direction of the {\it in situ} star AM. We find that the clearest observational consequence of this reorientation is that 81\% of halos rotate in a prograde direction, with a broad correlation between the rotation velocity and the dominant merger's infalling orbital AM. This prograde halo rotation is a clear and testable signature of this important mode of angular momentum acquisition by galaxies. 

Our highlighting of the importance of mergers in the angular momentum evolution of galaxies is consistent with a wide variety of existing works. Disk tilting is a common response to a satellite merger, both in non-cosmological \citep[e.g.,][]{HuangCarlberg1997,Kaz09,Dodge2023} and cosmological simulations (\citealt{Welker2014}, \citealt{Earp2019}, \citealt{Dillamore2022}; see also N.\ Ash et al., in preparation). This study explores the largest sample of MW-mass galaxies studied to date (471 systems), allowing a substantially more precise prediction that 80$\pm2$\% of galaxies should show alignment between {\it in situ} star AM and the orbital AM of the infalling dominant merger, in agreement with the 12/15 disk-dominanted systems studied by \citet{Dillamore2022}. In addition to confirming the importance of merger-driven galaxy reorientation for disks, this work shows that  reorientation is a common phenomenon even among systems with less prominent disks. The most massive merger produces the most important effect (\citealt{Gomez2017}, \citealt{Dillamore2022}, see our \S \ref{sec:others}). In common with \citet{Earp2019}, \citet{Gomez2017} and \citet{Dillamore2022}, we find that the largest reorientations tend to occur for the systems with the largest initial misalignments. We find that this relationship is very scattered, where only 35\% of the variance is accounted for by this relationship between initial orientation and the change in specific AM. Furthermore, our analysis shows that galaxies with high specific AM today tend to be those that have experienced the most significant merger-induced reorientation, typically through late-stage mergers with substantial satellites. To our knowledge, this provides the first systematic evidence from a large cosmological sample that high-AM galaxies can be the products of significant merger-driven reorientation. 

Turning to stellar halos, our work confirms and extends previous results. In common with \citet{Gomez2017}, \citet{Santistevan} and \citet{Beltrand2024}, we find that most stellar halos show prograde motions. By virtue of the size of the TNG-50 sample, we can quantify the expected fraction of prograde halos with more precision than in previous work to be 81$\pm2$\% (compared to e.g., 11/12 halos from \citealt{Santistevan}). We extend this result by pointing out the connection between merger-induced disk tilting and this tendency towards prograde stellar halo kinematics. This gives a qualitatively and quantitatively new probe of disk angular momentum acquisition suitable for study with powerful IFU and multiplexed spectrographs with 10-meter and 30-meter class telescopes. 

Because some of the alignment that happens between disks and infalling satellites happens as the satellites fall in, one might imagine (weaker) connections between the orbits of satellites and the angular momentum of the present-day main galaxy --- so-called spin--orbit alignment. Our above analysis (Fig.\ \ref{fig:dir_AM}, pink dashed lines) suggest that there should be very little detectable correlation between disk orientation and satellite AM at the time of infall. \citet{Moon2021} use TNG-100 to study spin--orbit alignment, the alignment between the spin of the main galaxy and the orbit of its nearest large satellite (for pairs with stellar mass ratios less than $\times$10 different from each other), finding that 56$\pm$1\% of pairs have prograde spin--orbit alignment (for gas particles), in agreement with the sign of our spin--dominant merger alignment. In addition, they find that the closest pairs have stronger alignment, up to $\sim70$\% prograde fraction for the closest pairs, qualitatively consistent with our findings. They find also that the strength of the alignment increases with the amount of time that a satellite has been influencing a central galaxy, which they interpret as favoring the idea that the central galaxy and its satellite are coming into alignment, rather than the spin--orbit alignment being a consequence of e.g., initial tidal torques from large scale structure. Our paper --- by showing little relationship between the spin and orbit at infall, but an increasingly close alignment of {\it in situ} stars during the interaction and merger process --- suggests that a major mechanism of spin--orbit alignment is the gradual tilting of the galaxy's spin towards the orbit of the satellite.    

Observationally, the results for spin--orbit alignment are a little mixed, but hint at signatures of a (weak) connection between the orbits of present-day satellites and disks \citep[e.g.,][]{Zaritsky1993,Hwang2010}. The later papers are most constraining because they draw from large samples, and generally find that the prograde fraction for satellites is within a few percent of 50\% \citep[e.g.,][]{Zaritzky1997,Azzaro2006,Herbert-Fort2008,Hwang2010}, very different from the $\sim80$\% prograde stellar halo fraction. Despite this near parity in the prograde--retrograde fraction, there are claims that the velocity distribution is asymmetric, with a mean velocity signifying prograde movement with $\sim35\pm3$\,km/s \citep{Herbert-Fort2008}, matching the typical halo rotation velocity in our sample. 

Merger-driven reorientation also carries subtle but observable implications for alignment between a galaxy's AM and the surrounding large-scale structure. As emphasized by \citet{Codis2012} and \citet{Welker2014}, because galaxy mergers tend to happen along filaments, their initial orbital AM tends to be oriented (roughly) perpendicular to the filament axis. This leads to a subtle predicted excess in AM orientation perpendicular to the filament direction for galaxies that have experienced the most merging \citep{Welker2014,GV2019}. While direct measures of merger history are lacking for large samples of galaxies, this expectation aligns with the observed tendency of elliptical and lenticular galaxies \citep{Tempel2013a,Tempel2013}, red galaxies \citep{Samuroff2023}, and galaxies with large bulges \citep{Barsanti2022} to possess AM perpendicular to their host filaments. 

Considering for a moment the MW specifically, while a number of papers have explored the effects of the MW's dominant merger on the orientation of the Galactic disk \citep[e.g.,][]{Dillamore2022,Dodge2023}, we suggest on the basis of stellar halo kinematics that the Milky Way may be an exception to this broad expectation. The Gaia-Enceladus/Sausage (GES) merger debris suggests a largely radial, low angular momentum orbit with minimal (and perhaps retrograde) rotation in the direction of disk rotation \citep{Helmi_2018,Belukorov_2018,Chandra2023}. In both this work and \citet{Dillamore2022}, it is emphasized that this mainly constrains the GES orbital angular momentum compared to the present-day disk orientation (as the original disk orientation is close to uncorrelated with the infalling satellite's orbital angular momentum). The GES merger appears to have been relatively early in the growth history of the Milky Way ($\sim$9\,Gyr ago; e.g., \citealt{Belukorov_2018,Gallart2019}). In TNG-50, such early mergers are less likely to affect the present-day orientation of the galactic disk (Figs.\ \ref{fig:delta_AM} and \ref{fig:halo_kin}). We therefore tenatively suggest that the present-day Galactic disk orientation, and the current tilt of the disk relative to the major axes of both the inner stellar halo \citep{Han2022} and DM halo \citep{Nibauer2025_dm}, may instead primarily reflect the accumulated effects of other important phenomena --- e.g., the angular momentum content of accreted gas \citep[e.g.,][]{Roskar2010,Debattista2015,Earp2019}, or torques from the DM halo \citep[e.g.,][]{Yurin2015,Debattista2015}. 

While stellar halo rotation is a powerful statistical test of galaxy reorientation by mergers, other analyses or diagnostics may be insightful for studying reorientation on a case-by-case basis. Disk warps may be one such probe (\citealt{HuangCarlberg1997}, \citealt{Kaz09}, \citealt{Deng2026}; however, see also \citealt{Sellwood2022}), including differences in warp signatures between different stellar populations \citep{Thul2025}. In the case of the MW particularly, detailed study of the warping and tilting of the MW's disk in response to the LMC may be quite informative (e.g., \citealt{Laporte2018}, N. Ash et al., in preparation). M31 may be another powerful test case. Wide-area kinematic measurements of M31's stellar halo (e.g., \citealt{Ibata2005}, \citealt{Gilbert2018}, \citealt{Escala2023}, \citealt{Dey2023}) show a complex velocity field, replete with substructure and showing clear prograde stellar halo motions, linked with M31's relatively recent massive merger \citep[last 2--4\,Gyr; ][]{Fardal2012,Hammer_2018,DSouza2018_M31,Dey2023}. Such young halos are precisely those where kinematic measurements are able to most powerfully inform our understanding of the progenitor's orbit (Fig.\ \ref{fig:halo_kin}). Furthermore, M31 has a very high specific AM $|\vec j_{\mathrm{M31}}| \sim 2 r_h v_c \sim 2 \times 5.3$\,kpc$ \times 255$\,km\,s$^{-1} \sim 2700$\,kpc\,km\,s$^{-1}$ \citep{Corbelli2010,Courteau2011}; galaxies with high specific AM are often those that have reoriented the most (panel a of Fig.\ \ref{fig:delta_AM}). It may be particularly useful to attempt to understand the warp of M31's disk \citep{Braun1991,Ferguson2002,Corbelli2010} as a possible observational diagnostic of the degree of merger-induced reorientation that it may have experienced.  

This work leaves a number of questions unaddressed. No attempt has been made to compare with resolved star observations of stellar halos --- for MW-like disk galaxies, such measurements are extremely sparse, owing to the difficulty of the observations. Furthermore, there are dramatic differences in the nature of the observations from galaxy to galaxy, from 6-D phase space information for the MW (e.g., \citealt{Helmi_2018}, \citealt{Belukorov_2018}) to wide-area kinematics for M31 with complex spatial and stellar population selection functions (e.g., \citealt{Ibata2005}, \citealt{Gilbert2018}, \citealt{Escala2023}, \citealt{Dey2023}) to a 1kpc$^2$ pencil beam in the stellar halo of MW analog NGC 4945 \citep{Beltrand2024} to a $\sim 20\times20$\,kpc multi-object spectrograph survey of NGC 253's inner stellar halo (K.\ Gozman et al., in preparation). Notwithstanding these differences, it is intriguing that of three halos with published resolved star kinematic measurements, only M31's stellar halo has significant prograde rotation, while the MW and NGC 4945 have little (and possibly retrograde) rotation \citep{Ibata2005,Dey2023,Helmi_2018,Beltrand2024}. 

Other possible paths to accreted star kinematics exist and may be more fruitful. Both globular clusters \citep[e.g.,][]{Brodie2014,Forbes2017,Li2026} and planetary nebulae \citep[e.g.,][]{Romanowsky03,Coccato} are potentially very fruitful as rare but bright kinematic tracers, particularly in the low surface brightness outer parts of elliptical and lenticular galaxies \citep[e.g.,][]{Pulsoni2018, Dolfi2021,Pulsoni2023,Bhattacharya2023}.
Substantial samples have now been assembled and frequently show kinematic transitions towards the galaxies' outskirts, often showing rotation \citep{Coccato,Dolfi2021,Pulsoni2018}. For fast-rotating ellipticals and lenticulars, the central and halo kinematic axes often align; in contrast, large misalignments in rotation direction are usually seen in many slow rotators \citep[e.g.,][]{Pulsoni2018}. Some of these results have been carefully compared with expectations from the TNG-100 simulation, showing agreement with the AM content expected of ellipticals and lenticular galaxies \citep{Pulsoni2023}. An extension of this TNG-50 analysis effort focusing on early-type galaxies and the relevant observables may be the best practical prospect for bringing to bear a substantial observational dataset on the problem of galaxy AM acquisition and reorientation, albeit in systems with typically lower AM and more complex merger histories. 

On the modeling side, there is scope to extend the type of insightful thinking characteristic of the work by \citet{Dodge2023} and others to systems with cosmological accretion and growth histories (see N.\ Ash in preparation). Most of the variance in the AM change from the accretion time to the present day was not accounted for by any of the accretion parameters that we tested. Gas accretion and DM halo both affect galaxy orientation \citep{Debattista2015}; it will be important to understand how to simplify one's thinking about cosmological simulations to tease out how each of the ingredients combine to impact galaxy orientation (perhaps by using `genetic modification' to vary merger history in more controlled ways; e.g., \citealt{Rey2023}). An additional promise of such an analysis is that it may be able to understand if other facets of a galaxy --- e.g., its structure or its SFH --- are influenced specifically by reorientation in a way that can be tested using stellar halo kinematic measurements. 

\section{Summary}

This work explores the response of the AM of the main bodies of galaxies to their hierarchical growth histories. Using a sample of 471 TNG-50 galaxies, we find that the main bodies of galaxies often reorient in response to mergers with a large satellite, with approximately 80$\pm2$\% of galaxies aligning their present-day AM towards the orbital AM of their most massive (dominant) merger progenitor; other satellites have a small to negligible influence. This realignment typically involves a median change of $\sim$50\% in the galaxies' specific angular momentum, a shift that is most pronounced in galaxies with the largest initial misalignments. However, this reorientation is rarely total or perfect; the median misalignment remains approximately 46{\degr}, underscoring that while mergers are an important driver, they combine with other factors such as DM halo triaxiality and misaligned gas accretion to determine a galaxy's ultimate orientation. 

The most accessible observational record of reorientation by mergers is the accreted stellar halo. We show that halo rotation serves as a testable signature of galaxy reorientation, as both components tend to dynamically couple and align towards the orbital vector of the dominant merger. Consequently, 81$\pm2$\% of TNG-50 stellar halos exhibit prograde rotation relative to the main body of a galaxy, a feature that should be detectable with future high-precision kinematic surveys. Halo kinematic measurements for systems with more recent mergers ($\lesssim7$\,Gyr ago) are predicted to be particularly informative, with the degree of halo rotation broadly correlating with the specific AM of the dominant merger. Conversely, the orbital characteristics of ancient accretions are significantly harder to recover, a category that likely includes the Milky Way. Because its dominant merger occurred early and on a nearly radial orbit, the Galaxy’s present-day orientation may instead reflect accumulated effects from gas accretion or DM halo torques. 

\vspace{-0.3cm}
\begin{acknowledgments}

We thank the referee for their useful comments and helpful suggestions. We appreciate useful conversations with and suggestions from Neil Ash, Brigette Vazquez Segovia, and Yingtian (Bill) Chen. 
This work was supported in part by NASA under grant number 80NSSC24K0084 under solicitation NNH22ZDA001N-ROMAN: Nancy Grace Roman Space Telescope Research and Support Participation Opportunities.

\end{acknowledgments}

\paragraph{Software}
 \texttt{Matplotlib} \citep{matplotlib}, \texttt{NumPy} \citep{numpy-guide,numpy}, \texttt{Astropy} \citep{Astropy13,astropy} %\texttt{SciPy} \citep{scipy},

%% For this sample we use BibTeX plus aasjournalv7.bst to generate the

\bibliography{diskAM}
\bibliographystyle{aasjournal}

%\bibliographystyle{apj}



\end{document}